\documentclass[aps,pre,preprint,superscriptaddress,showpacs]{revtex4}
\usepackage{epsf}

\begin{document}

\title{Non-Gaussian PDFs from Maximum-Entropy-principle considerations}

\author{Fabio Sattin}
\email{fabio.sattin@igi.cnr.it}

\affiliation{Consorzio RFX, Associazione Euratom-ENEA, Corso Stati Uniti 4, 
35127 Padova, Italy}

\begin{abstract}
In this work we develop on the recently suggested concept of
superstatistics [C. Beck and E.G.D. Cohen, Physica A {\bf 322}, 267 (2003)],
face the problem of devising a viable way for estimating the correct
statistics for a system in absence of sufficient knowledge
of its microscopical dynamics, and suggest to solve it through the Maximum 
Entropy Principle. As an example, we deduce the Probability
Distribution Function for velocity fluctuations in turbulent fluids,
which is slightly different from the form suggested in [C. Beck, Phys. Rev. Lett. 
{\bf 87}, 180601 (2001)].
\end{abstract}


\maketitle

Nonextensive statistical mechanics (NESM) has been raising considerable 
interest along these last 15 years. However, while there is an universal 
consensus about its most famous prediction, {\it i.e.} the ubiquitous 
existence in nature of power-law Probability Distribution Functions (PDFs), the
soundness of the theoretical foundation on which is based-the generalized definition 
of entropy \cite{tsallis}-has been questioned by several authors. 
We quote here, e.g., the paper \cite{nauenberg} which dealt in particular 
with contradictions between theory's predictions and thermodynamical 
constraints.\\
The paper \cite{sattinepl} started from these critiques but focussed 
more on the information--theoretic aspect, suggesting that Tsallis' 
entropy should be regarded merely as a practical tool for doing predictions in 
presence of a reduced amount of information about the system. Indeed, the need
of resorting to modified definitions of the informational entropy 
when the knowledge about the states of the system is insufficient, is
well known in statistics and has been extensively pointed out in 
\cite{luzzi1} and references therein. \\ 
Quite recently, Sattin and Salasnich \cite{pre}--starting from
an earlier work by Beck \cite{beckprl}--and, in a more formal 
and rigorous way, Beck and Cohen \cite{superstatistics} 
demonstrated, without direct reference to any definition of
entropy, that Tsallis' statistics is just a particular case
of an infinite class of statistics (hence the name 
``superstatistics''). All the elements of the class are characterized by one or
more parameters, and reduce to ordinary Maxwell-Boltzmann 
statistics for particular values of the parameters. Beck and Cohen 
started from a model of dynamical system, a Brownian particle moving 
according to a Langevin equation, where the noise ($\sigma$) 
and friction ($\gamma$) terms
are allowed to fluctuate, to arrive to the famous equation
\begin{equation}
    \label{eq:pe}
    P(E) = K \int_{0}^{\infty} e^{-\beta E} f(\beta) d\;\beta
\end{equation}
Here, $P(E)$ is the PDF for the system of being in the state of 
``energy'' $E$, $\beta$ is a fluctuating parameter (generalized 
inverse temperature), which in the 
present formulation is function of $\sigma$, $\gamma$, and
$f(\beta)$ is the PDF for the realization of the particular value 
$\beta$; $e^{-\beta E}$ is the usual Boltzmann factor. 
Eq. (1) is more general than suggested by the Brownian 
particle model. Indeed, it can be written for any system $\Sigma$
interacting with a fluctuating environment $B$, hence $P(E)$ is a weighted average 
of the standard Boltzmann statistics over different realizations of
the interaction between the system and its environment, quantified by
$f(\beta)$. Notice that the existence of finite fluctuations 
in the environment is strictly correlated with finite-size-effects
of the environment itself; indeed, the result that Tsallis statistics
can arise within the framework of a system interacting with
a finite thermal bath was recently reported by Aringazin and Mazhitov 
\cite{aringazin}, and it had already been suggested much earlier 
by Plastino and Plastino \cite{plastino}.\\
The physical content of the theory, thus, shifts from the entropic 
index $q$ to the PDF $f(\beta)$. Of course, the explicit expression 
for $f(\beta)$ must vary for any single problem, constrained just by some
rather intuitive criteria (normalizability, etc \ldots); 
Beck and Cohen give several possible 
examples of functions which are potential candidate for $f(\beta)$. 
However, a simple criterion able to guide the user towards a plausible 
functional form for $f$, lacking a more detailed knowledge of the
underlying microscopical details, would be very satisfying. But such a
criterion is readily available: it is the well known Jaynes' 
Maximum Entropy  Principle. In this case, the system about which we
have not a proper knowledge is no longer $\Sigma$ but the environment
$B$ itself, and $f(\beta)$ is a measure of the probability of $B$ of  
occupying a state in an abstract one-dimensional space parameterized
by $\beta$. \\
Lacking any further information, the 
most probable realization of $f(\beta)$ will be the one that maximizes 
(Shannon) entropy  $S(f) = - \int f \ln f d\beta$ with suitable 
constraints. \\
In this paper we present a straightforward application of this 
principle to an important example, namely the PDF of velocity 
fluctuations in turbulent flows \cite{bls}. This accurate 
experimental measurement is thought to be one of the strongest 
evidences in support of Tsallis' theory, since the empirical PDF
appear very well matched by power-laws. In this work we suggest 
instead that the true curve can be very close numerically, but rather
different in its analytical expression, from a power-law. Indeed, 
further experimental investigations of turbulent flows are now
suggesting the existence of small deviation from pure power-laws
(experiments of Jung and Swinney, cited in \cite{superstatistics}). 
Recent work by Aringanzin and Mazhitov \cite{ar1} deal with
the attempt of theoretically recovering the new accurate experimental 
distributions for fluid particle accelerations. In this work,
instead, we will consider, only fluid velocity differences. \\
As a first test case, let us suppose of knowing anything about $f$
but the average value of the inverse temperature 
\begin{equation}
<\beta> = \int \beta f(\beta) d\beta \quad .
\end{equation}
The most probable $f$ extremizes the functional
\begin{equation}
F = - \int f \ln f d\beta - \lambda \int f \beta d\beta \quad .
\end{equation}
and the solution is the Boltzmann-like function
\begin{equation}
    \label{eq:f0}
f(\beta) = f_{0} \exp(- \lambda \beta)
\end{equation}
($\lambda$ is the Lagrange multiplier). Such an expression was 
obtained, by example, in \cite{batanov}, from which the original
idea of entropy maximization was taken. It reads slightly different
from the Gamma (or $\chi^{2}$) distribution originally devised by Beck \cite{beckprl}
to recover power-law PDF. If we replace $f$ of (\ref{eq:f0}) into
(\ref{eq:pe}) we get
\begin{equation}
    P(E) \propto {1 \over E + \lambda} \quad ,
\end{equation}
which is not a physically acceptable solution, since $P(E)$ must be normalizable.\\
In order to make a step beyond this rough scheme it is necessary to 
input more information within the model. We do it starting from the 
same premises as Beck's \cite{beckprl} but diverging at just the next step: 
infact, he defines
$\beta \propto \epsilon_{r} \tau$, where $\epsilon_r$ 
is the energy dissipation rate of the fluid on microscopical scale, 
and $\tau$ the typical energy transfer time. We note 
that $\epsilon_{r} \tau $ has units of [energy], not of [energy]${}^{-1}$, 
and a more intuitive way of writing this relation should be 
 $\beta \approx (\epsilon_{r} \tau)^{-1}$. However, the inverse 
of a sum of squares of random variables does not yield a 
$\chi^{2}$-distributed random variable, hence Beck is 
forced to re-establish the correct dimensions by multiplying by the 
constant $\Lambda$, with the units of [speed]${}^{4}$. 
Thus, a characteristic speed $\Lambda^{1/4}$ has entered the calculations, whose 
physical interpretation remains obscure. Of course, Beck was forced
to do this assumption because of the sought agreement with 
Tsallis' theory. Now, we are free from this constraint, and can 
allow for more natural choices, although we must agree that a certain arbitrariness 
in the choice of the physically meaningful variables is unavoidable within 
this context. \\
To start with, we shift from $\beta$ parameter to $T = 1/\beta$. $T$, 
which has units of [energy] is a more convenient variable, as explained 
above. We can write the equivalent of Eq. (1) in terms of the new 
parameter; furthermore, in order to adhere to existing literature, 
it is convenient using a generalized velocity instead of energy:
$ E = u^{2}/2$. Therefore Eq. (1) becomes
\begin{equation}
\label{eq:pu}    
p(u) = \int dT \sqrt{1 \over 2 \pi T}  \exp\left(- { u^{2} \over 2 
T}\right) g(T) \quad .
\end{equation}
Again, we follow Beck's recipe: 
the parameter $T$ is written in terms of fluctuating Kolgomorov 
velocities $u_{i}$
\begin{equation}
\label{eq:t}    
T = {T_{0} \over 3} \sum_{i=1,2,3} u_{i}^{2}
\end{equation}
where $T_{0}$ is a constant. We stress that the r.h.s. of the 
previous equation and that of Eq. (28) in \cite{beckprl} are the same, 
although the l.h.s.'s are each the inverse of the other. This is due to 
our discarding of the artificial constant $\Lambda$. $T_{0}$, 
conversely, has a straightforward physical interpretation as average thermal 
energy. \\
The PDF $g(T)$ can be rewritten as a function of $u$'s: 
$ g'(u_{1},u_{2},u_{3}) \equiv g'({\bf u}) \leftrightarrow  g(T) $. 
The Maximum-Entropy principle imposes of extremizing
\begin{equation}
F =  - g'({\bf u}) \ln( g'({\bf u}))  - {1 \over 
u_{M}^{2}} (u_{1}^{2} + u_{2}^{2} + u_{3}^{2}) g'({\bf u})  
\end{equation}
where $1/u_M^{2}$ is the Lagrange multiplier corresponding to the fact 
that we fix the average kinetic energy. The solution is, obviously,
$g'({\bf u}) = C\exp(-(u_{1}^{2}+u_{2}^2+u_{3}^{2})/u_{0}^{2})$. From 
here, reversing to $T$ variable,
\begin{equation}
\label{eq:gt}    
g(T) = K T^{1/2} \exp\left(-{3 T \over 2 T_{0}}\right) \quad . 
\end{equation}
The term $T^{1/2}$ comes from the volume element, $C, K$ are normalization 
constants and we have rescaled velocities such that $u_{M}^{2} \equiv 2/3$. 
Notice that Eq. (\ref{eq:gt}) is a Gamma distribution and 
could be obtained straightforwardly from Eq. (\ref{eq:t}) by assuming 
from the start that the $u_{i}$ were normal random variables, just as done 
by Beck. \\
Let us now replace Eq. (\ref{eq:gt}) into (\ref{eq:pu}): we get
\begin{equation}
p(u) = K' \int_{0}^{\infty} \exp\left(- {u^{2} \over 2 
T} - {3 T \over 2 T_{0}}\right) \, dT \quad .
\end{equation}
The explicitly normalized solution reads
\begin{equation}
    \label{eq:ppw}
p(u)_{PW} = {1 \over \hat{u} \pi} {u \over \hat{u}} K_{1}\left( { u \over 
\hat{u}} \right) \quad \left(\hat{u} = \sqrt{ T_{0} \over 3} 
\right)     
\end{equation}
and $K_{1}$ is the Bessel $K$ function of order one.
This result appears rather different from usual power-laws. Indeed, we 
chose to plot in Fig. (1) this curve together with the best fitting curve found by 
Beck for the velocity PDF:
\begin{equation}
    \label{eq:pbeck}
  p(u)_{\text{Beck}} = {1 \over Z_{q}} {1 \over \left( 1 + (q - 1) \tilde{\beta} C |u|^{2 
  \alpha}  \right)^{1/(q-1)} } \; (q \approx 1.1 , \alpha \approx 0.9).
\end{equation}  
(See dashed line in Fig. 1 of \cite{beckprl}). On the whole, the two 
curves match rather closely. Some differences appear at low $u$'s.

\begin{figure}
\epsfxsize=12cm
\epsfbox{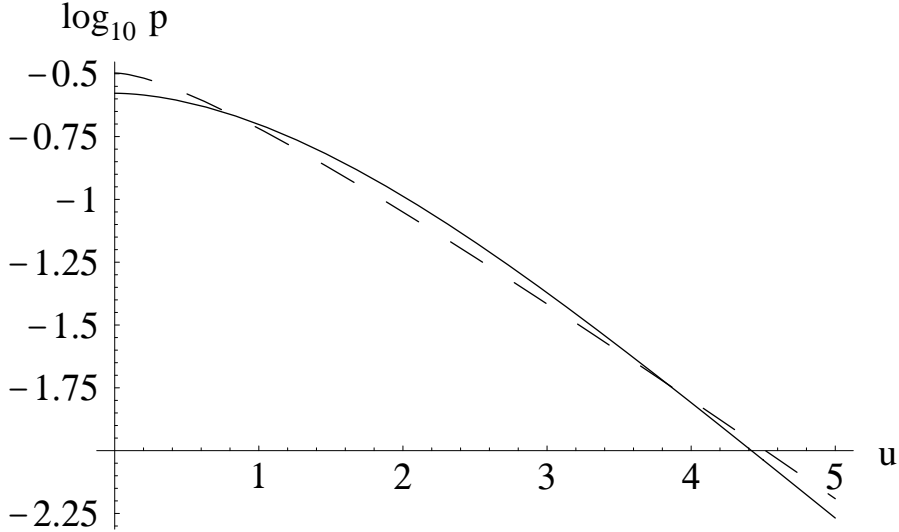}
\caption{Solid line, $p(u)_{\text{Beck}}$ from Eq. (\ref{eq:pbeck}); 
dashed line, $p(u)_{PW}$ from present work (Eq. \ref{eq:ppw}). 
The adjustable parameters have been chosen 
so that the curves have the same variance.}
\label{fig:uno}
\end{figure}

Since we have not available the original experimental data $p_{\text{expt}}$, 
we cannot directly compare on them the goodness of our fit. A good insight 
comes, however, by plotting the relative difference $\Delta = 
(p_{\text{Beck}} - p_{PW})/p_{\text{Beck}}$ (Fig. 2, $p_{PW}$ is our 
solution, given in Eq. \ref{eq:ppw}). The quantity 
$\Delta_{1} = (p_{\text{Beck}} - p_{\text{expt}})/p_{\text{Beck}}$ appears plotted 
in Fig. 2 of \cite{bls}. If $\Delta$ and $\Delta_{1}$ agree, $p_{PW} = 
p_{\text{expt}}$. It appears that our fit slightly overestimates experiment at $ u 
\approx 0$, but for the same amount as $p_{\text{Beck}}$ does 
underestimate it. On the whole, the agreement with experiment is remarkable.
We remark that, as only adjustable parameter, we used the 
hypothesis that $T$ is of the form (\ref{eq:t}) with the index ranging 
from one to three. The equivalent of $\alpha$ parameter used by Beck does not enter
our calculations. Thus, we have realized an economy in our way of 
modelling the data.

\begin{figure}
\epsfxsize=12cm
\epsfbox{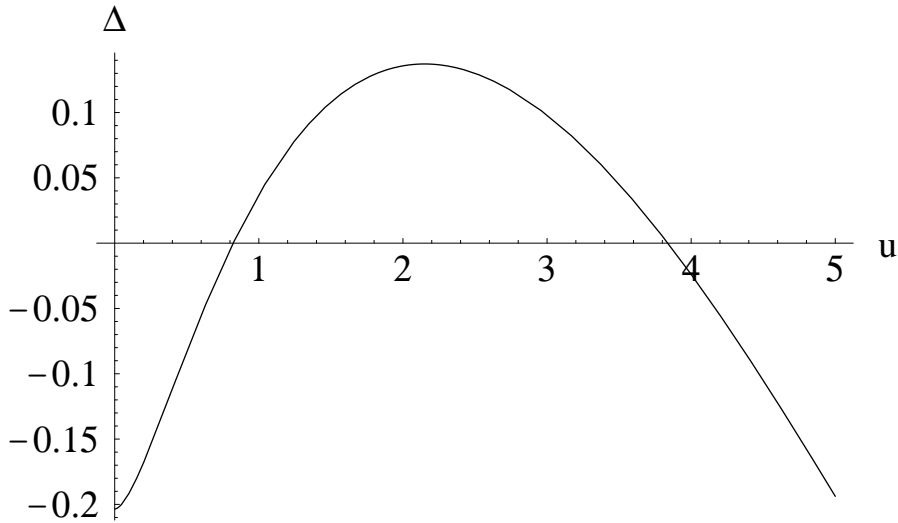}
\caption{The relative difference of the two curves above: $\Delta = 
(p_{\text{Beck}} - p_{PW})/p_{\text{Beck}}$ versus $u$, where $p_{\text{Beck}}$ 
is from Eq. (\ref{eq:pbeck})) and $p_{PW}$ is from Eq. (\ref{eq:ppw}). 
This figure should be compared with Fig. 2 of ref. \cite{bls}.}
\label{fig:due}
\end{figure}

We call the reader's attention to the 
fact that, in his Fig. 1, 
Beck studied at once (i) velocity spatial differences, and (ii) 
accelerations. We, instead, considered only the former quantity, and it is 
possible to see that the latter cannot be reproduced by the present 
treatment, even allowing for a varying number of Kolmogorov velocities 
in Eq. (\ref{eq:t}). This must be traced back to the fact 
that now $g(T)$ is no longer a good weight function, but we must define 
an equivalent of $T$, defining the ``average acceleration'' of the system.  \\
Finally, we point to two important issues: 
I) we have the asymptotic trend $x K_{1}(x) \to x^{1/2}\exp(-x), \; (x \to \infty)$, 
that is, we have not power-law decay. Notwithstanding 
this, our curve nicely fits data that have been previously considered as 
stemming from a power-law PDF. The reason lies in the finite $u$-range 
sampled and therefore calls for an inherent ambiguity in this kind of 
studies, related to finite experimental scans: unless one is sure of 
being investigating the true asymptotic region, one can never be
completely confident about the fitting curve used. \\
II) As reported above, all superstatistics must collapse to the single 
Boltzmann statistics when $ q \to 1$, that is, differences between
different models are at least order $ {\cal O}(q -1)$. Since, in this case, we are 
dealing with a parameter not far from unity, $ q \approx 1.1$, it is 
to be expected that any two reasonable models would give close results.
(The author wishes to thank C. Beck for calling these two points to his 
attention).

In conclusion, the use of the concept of superstatistics together 
with Maximum-Entropy principle appears to be an efficient way 
of estimating statistical properties in general systems using a
minimal amount of information.

\begin{acknowledgments}
The author wishes to thank A. Rapisarda, C. Beck, and R. Luzzi for useful comments and 
suggestions.    
\end{acknowledgments}

\end{document}